\documentclass[USenglish,11pt ]{article}

\usepackage{makeidx}         
\usepackage{graphicx}        
\usepackage{multicol}        
\usepackage[latin1]{inputenc}
\usepackage{palatino}
\usepackage{natbib}
\usepackage[top=2.4cm,left=3.4cm,bottom=2.4cm,right=3.4cm]{geometry}

\usepackage{amsmath}
\usepackage{amssymb}
\usepackage{enumerate}

\usepackage{hyperref}
\hypersetup{colorlinks,
            citecolor=black,
            filecolor=black,
            linkcolor=black,
            urlcolor=black}

\newcommand{\Proxm}{P_p}
\begin{document}
\title{Science mapping with asymmetrical paradigmatic proximity}
\author{Jean-Philippe Cointet\footnote{Both authors have equally contributed to this work}\,$^{,}$\footnote{CREA (Center for Research in Applied Epistemology), CNRS/Ecole Polytechnique, 1 rue Descartes, 75005 Paris, France.}\,$^{,}$\footnote{ TSV (Social and Political Transformations related to Life Sciences and Life Forms), INRA, 65 Boulevard de Brande-
bourg, 94205 Ivry-sur-Seine Cedex France.}
 \& David Chavalarias\footnotemark[1]\,$^{,}$\footnotemark[2]\,$^{,}$\footnote{ISCPIF (Complex Systems Institute, Paris Ile-de-France), 57-59 rue Lhomond, 75005 Paris, France.}
 \vspace{0.2cm}\\ {\small \em 
      \{jean-philippe.cointet,david.chavalarias\}@polytechnique.edu}}

\maketitle 

\begin{abstract}
We  propose a series of methods to represent the
evolution of a field of science at different levels: namely micro, meso and macro levels. We use a previously introduced asymmetric measure of paradigmatic proximity between terms that enables us to extract structure from a large publications database. We apply
our set of methods on a case study from the complex systems community through the mapping of more than $400$ \emph{complex systems science concepts} indexed from a database as large as
several millions of journal papers. We will first summarize the main properties of our asymmetric
proximity measure. Then we show how salient paradigmatic fields can be embedded into a
2-dimensional visualization into which the terms are plotted according to their relative specificity and generality index. This meso-level helps us producing macroscopic maps of the field of science studied featuring the former paradigmatic fields.
\end{abstract}

\section{Introduction}

Scientific activity can be interpreted as a complex process \cite{Hull-1988} that derives from a large scale interaction
network made of heterogeneous actors: scientists, engineers, natural objects, journals,
public and private laboratories\cite{Latour:act}, etc...   The main core of this large scale intertwining system can be coarsely resumed to scientists publishing articles through new collaborations that synthesize a state of knowledge at a given time.
We consider that scientific publications are among the major products of scientific communities and that they make possible the coordination of millions of people distant in space and time  \cite{latour:cons}. As such publications are one of the main communication
medium for scientists. This is a stigmergic media since it is not directed toward a
specific recipient, information are public and each new associations between concepts shall modify even slightly the scientific landscape.

	\begin{figure}
	\centering
 	\includegraphics[width=0.6\textwidth]{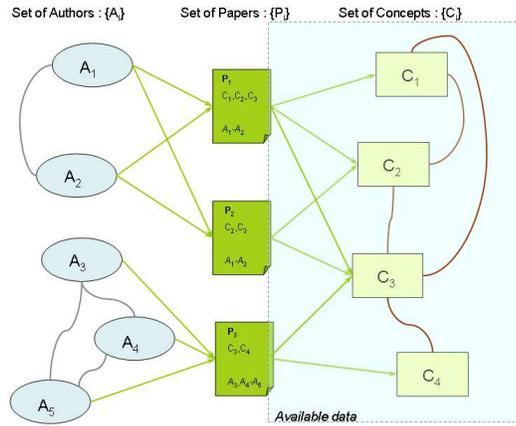}
 	\caption{\textbf{Scientific knowledge production scheme}: a set of authors $\{A_i\}$ produce publications $\{P_i\}$ which introduce concepts $\{C_i\}$. We defined paradigmatic field as strongly cooccurring set of concepts.}\label{scheme1}
 	\end{figure}

The figure \ref{scheme1} represents the scientific knowledge production process. Scientists $\{A_i\}$ publish papers $\{P_i\}$ that introduce a set of concepts $\{C_i\}$ and their relations.
Following Kuhn's observation that ``a paradigm is what the members of a scientific community share, and,
conversely, a scientific community consists of men who share a paradigm'' \cite{Kuhn:1970a} we make the assumption that there is a strong correlation between the structures of the scientific communities (on the left of the diagram) and the structures of terms co-occurrences (on the right) that represent inner constituents of the different paradigms and their articulation.

\begin{figure}
 	\centering
 	\includegraphics[width=0.4\textwidth]{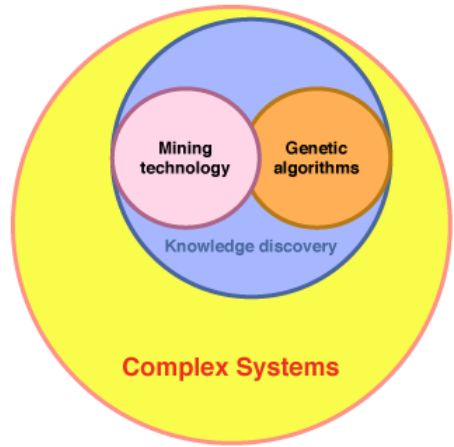}
 	\caption{\textbf{Schematic example of organization of paradigmatic field}: Typical concepts organization. Concepts area corresponds to their occurrences and overlapping areas to co-occurring concepts.}
 	\label{scheme2}
 	\end{figure}

We shall define a paradigmatic field as a set of concepts that reflects the structure of the activity of scientific communities. We are then looking for characteristic patterns of words co-occurrences corresponding for example to hierarchical structures (in our schematic example figure \ref{scheme2} \emph{knowledge discovery} is embedded in the \emph{complex systems} field). Terms may also be part of close though distinct paradigmatic fields (as illustrated figure \ref{scheme2}, \emph{knowledge discovery} may be used by machine learning community (\emph{genetic algorithm}) but also by data mining community (\emph{Mining technology})). We will thus have to develop overlapping paradigmatic fields detection.


The aim of this paper is to elaborate on the key benefits we derived from an asymmetric proximity measure previously introduced \cite{chava:scien}. We will then present methods and tools for automatic bottom-up identification of  multi-scale structure of paradigmatic
fields and apply these onto a case study concerning complex systems science.

This measure is based on mere statistics on occurrences and co-occurrences of terms use in a scientific database. In the first part of this paper, we shall explain the main properties of this proximity measure and show its advantages compared to other classical measures. We will then describe the clustering method we use to detect paradigmatic fields and define a two dimensional space that helps us representing them in an informative way. The last part of this paper will deal with high-level representation of a large scale corpus. We will finally propose a method to represent in an understandable way a large set of terms at the macro-level. 

\section{Context and rationale}

Retrieving structure from a set of terms occurrences and co-occurrences is one of the main objective of scientometry.
Doyle  was among the first to point to the fact that navigation in large scientific database was ineffective due to the lack of relevance of traditional document retrieval techniques \cite{doyle}. One method that has been proposed and largely commented is ``co-word analysis''. A classical statistics in co-word analysis
which has been extensively used
\cite{calsoc,callon91coword,Noyonsb} is the similarity index given by the ratio between the number of co-occurrences between
the two terms $a$ and $b$ divided by the product of the number of total
occurrences of $a$ and $b$.

This similarity index is entirely symmetric which means that given a term $a$ and another term $b$, $a$ is at the same distance from $b$ than $b$ from $a$. This \emph{a priori} symmetric constraint can happen to be problematic when comparing terms of different ``frequencies''. Let's consider a case where every occurrence of $b$ is followed by an occurrence of $a$ which may be the case if $b$ refers to the sub-field of a more generic field $a$ (for example \emph{mining technology} can be described as a subfield of \emph{knowledge discovery}). In this case, we would like to be able to derive this hierarchical relation directly from our proximity measure. This is impossible with the classical proximity index which will not allow to exhibit this kind of hierarchical structure.

In order to retrieve this kind of structure we proposed \cite{chava:scien} an alternative proximity measure between terms $i$ and $j$ which we called \textit{paradigmatic proximity} : $$\Proxm^\alpha(i,j)=(n_{ij}^t/n_{i}^t)^{\alpha}(n_{ij}^t/n_{j}^t)^{1/\alpha}$$
where $n_{i}^t$ and $n_{j}^t$ are the number of occurrences of $i$ (respectively $j$) at time $t$ and $n_{ij}^t$ is the number of co-occurrences of $i$ and $j$. The ``focus parameter'' $\alpha$ is a tunable real positive parameter. This paradigmatic proximity has the following properties (we do not mention the parameter $\alpha$ for sake of clarity):



\begin{enumerate}
\item $\Proxm(i,j)=0$ if $n_{ij}^t=0$
\item $lim_{\frac{n_{ij}^t}{n_i^t}\rightarrow 0}(\Proxm(i,j))=0$
\item $\Proxm(i,i)=1$
\item $\Proxm(i,j)$ is growing with $n_{ij}^t$ since larger co-occurring
sets illustrate higher paradigmatic proximity. On the contrary, growing $n_i^t$ or $n_j^t$, $n_{ij}^t$ being constant, will decrease the value of $\Proxm(i,j)$
\item Under the assumption that the sample is representative $\Proxm(i,j)$ is
independent of the total number of articles in the database.
\end{enumerate}

Our paradigmatic proximity enables to define the neighborhood of a
target term $i$ given a threshold $s$ and an $\alpha$ value at
time $t$ by :
$$V_{s,\alpha}^t(i)=\{j|\Proxm^{\alpha}(i,j,t)>s\}$$
As we shall see in section 4.1, tuning $\alpha$ enables us to describe the way a
term belongs to a sub-field of a target term or on the
contrary how a target term is part of more generic fields.

\section{Case study}
Our case study focuses on a set of terms coming
from two sources : a set of key-words associated to complex systems European projects extracted from IST Cordis database of FP6 and FP7
($765$ key-words generously provided by the Arc System team lead by
Joseph Fröhlish) and a set of key-words
collected near colleagues (about one hundred).
We established a
partnership with Scirus, Elsevier's free science-specific search
engine\footnote{\href{http://www.scirus.com}{http://www.scirus.com}} in order to collect the number of
occurrences and co-occurrences per year of these concepts from 1975
to 2005 in the full text of the articles. The database is made of more
than $20.000.000$ publications covering a wide range of
scientific content platforms\footnote{ BioMed Central,
Crystallography Journals Online, Institute of Physics Publishing,
MEDLINE/PubMed,
 Project Euclid, ScienceDirect, citation, Society for Ind. \& App. Mathematics and Pubmed Central}.

Due to the numerous access to the database, data collection was pretty slow and we had to
narrow our set of concepts to 448 terms\footnote{The set of terms can be consulted on  \href{http://isc-pif.csregistry.org/complex+systems+terms}{\textbf{http://isc-pif.csregistry.org/complex+systems+terms}}}. Since co-occurrences extraction was very demanding in terms of
server availability, we also decided to send a query for a
co-occurrence of two terms only when the two queries on single terms gave a non
zero result in the  ``authors key-words'' field (each concept has been
mentioned at least once as an article key-word for the
year considered). Consequently our database is made of all queries
results for single terms in full text from 1975 to 2005, and every
query results on full text co-occurrences for pairs of concepts
that both appeared at least once as author key-words the year
considered.

This rough statistics enables to compute the paradigmatic proximity for any
time window from 1975 to 2005. If we choose a time range between
years $Y1$ and $Y2$, we thus have the following extended formulation
of paradigmatic proximity given this time range:

$$\Proxm^{\alpha}(i,j,[Y_1 ...
Y_2])=(\frac{\sum_{t=Y_1...Y_2}n_{ij}^t}{\sum_{t=Y_1...Y_2}n_{i}^t})^{\alpha}
(\frac{\sum_{t=Y_1...Y_2}n_{ij}^t}{\sum_{t=Y_1...Y_2}n_{j}^t})^{\frac{1}{\alpha}}$$

We will now provide some examples of visualizations built upon our paradigmatic
proximity measure at different scales : micro, meso and macro levels. \footnote{It should not be forgotten that the set of terms describing the thematic fields that we will extract from our database shall necessarily be part of our 448 initial terms. Consequently some important terms may not appear in the following representations.}

\section{multi-level science mapping}

A classical objective in bibliometric literature is to draw knowledge maps \cite{buter,bibmap}.
Clustering methods like Kohonen maps algorithms have been used to
provide smarter navigation tools in articles databases thanks to
conceptual mapping of a wide research area \cite{Lin91,sunmed}.
Many approaches also propose to use both terms occurrences and
references to help producing knowledge maps \cite{gaston}.

Here our approach is restrained to the mere occurrences and co-occurrences statistics but we apply our asymmetric paradigmatic proximity in order to detect more detailed structure than classical flat maps from our set of terms as we are now able to distinguish between different levels of specificity/generality.

We propose to represent our initial set of concepts at three distinct levels of aggregation. First we  define a micro-level based on local neighborhoods defined by the proximity measure. Then meso-level enables us to define paradigmatic fields as consistent and relevant set of concepts. The macro-level built upon the former level provides an intelligible map of the complete scientific landscape.

\subsection{Micro-scale : paradigmatic neighborhoods}

\begin{figure}
  \centering
  \includegraphics[width=0.9\textwidth]{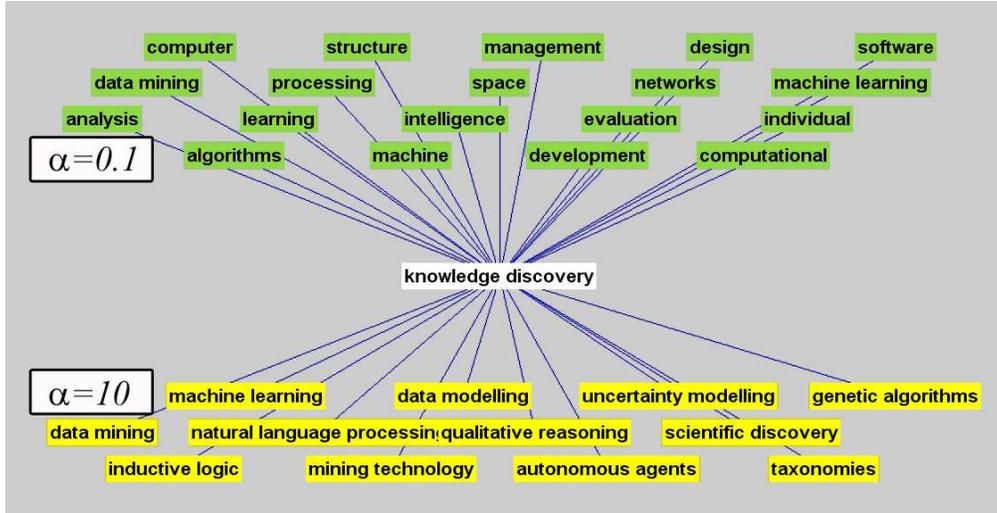}\
  \caption{\textbf{Specific and generic neighborhood of term \textit{knowledge discovery} given a threshold $s$}. bottom: $\alpha = 10$, the closest terms that
  specify \textit{Knowledge Discovery} ; top: $\alpha=0.1$,
   the closest terms that are more generic than \textit{Knowledge Discovery}.
   }\label{navig}
\end{figure}

Micro-scale is the most straightforward level.  Here the approach is local which means that we only refer to term-centered proximities. We simply look at the neighborhood of a target term $i$ with a given value of $\alpha$. This focus parameter enables to tune the required degree of specificity or generality of neighbors relatively to the target term. For low values of $\alpha$ ($\alpha<1$), the nearest neighbors tend to be more generic than the target term. For higher $\alpha$ ($\alpha>1$), we access to more specific terms.
The figure \ref{navig} illustrates this property around the target term \emph{knowledge discovery} extracted from the case study introduced in section 3. We plotted the neighborhood $V_{s,\alpha}$ of \emph{knowledge discovery} for $\alpha = 0.1$ and $\alpha = 10$ given a fixed threshold $s$. For larger $\alpha$ terms in the neighborhood of \emph{knowledge discovery} become
more specific and closer to the terms used by specialists of the
field. We thus get terms that sharply qualify areas of
investigations about knowledge discovery. On the contrary, values of $\alpha$ below $1$ will tend to reframe the target word within its broader context.

In the special case where $\alpha = 1$ we exclude very generic and very specific neighbors and our paradigmatic measure equals the equivalence index (e-coefficient) introduced by Callon \cite{callon91coword}.  Note that the property $\Proxm^{\alpha}(i,j)=\Proxm^{1/\alpha}(j,i)$ ensures that given a term $B$ at distance $\Proxm^{\alpha}(A,B)$ from a target term $A$, one can find $A$ in the neighborhood of $B$ exactly at the same distance ($\Proxm^{1/\alpha}(B,A)$) simply by switching $\alpha$ to $1/\alpha$. 

\subsection{Meso-scale: identification of paradigmatic fields}




 Looking at the bottom part of figure \ref{navig}, we can see that
 several distinct spheres of knowledge seem to co-exist that share the use of the term \emph{knowledge discovery}. One is ``machine learning'' oriented while the other is rather ``data mining'' oriented. Contextual information enables us to exhibit automatically these multiple practices.

Identifying set of
terms reflecting this various scientific practices requires a broader view of
the terms landscape structure taking into account every relations between
the different terms neighbors. We wish to automatically categorize our data according to the values of the paradigmatic proximity $\Proxm^{\alpha}$ given for each pair of terms. Since a term can be polysemic, it may be used by several scientific
  communities, the categorization algorithm should then be able to categorize a term in different clusters.
  The literature is rich with community detection algorithms in networks (for a review of recent methods and their comparison see \cite{Danon:2005p1591}), most of them aim at performing the best partition of nodes into communities. Here we are looking for overlapping communities. Several algorithms have recently been proposed to achieve such a task  ; one successful method in line with this requirement is the k-clique percolation
algorithm recently introduced by \cite{palla} that operates on graphs to
detect possibly overlapping communities.

We proceed by generating a lexical graph based on our
proximity measure by setting a threshold $s$ and linking each term
$i$ to its set of neighbors: $V_{s,\alpha}(i)$.
This enables to build a non-directed graph on terms.
Then we can apply the k-clique percolation algorithm which outlines
communities of terms that qualify distinct spheres of knowledge
production. The output of this algorithm is made of clusters of terms
such that within each cluster, one can perform a k-clique percolation (with $k\geq3$).
Clusters are a general property of the graph (though they may depend on $\alpha$ and $s$), they do not depend on a  predetermined target term.

The very latest enhancements of the k-clique percolation algorithm \cite{palla:wei,palla:dir} propose an extension of the algorithm to directed and weighted networks. This would allow us to treat directly our proximity matrix without having to define a threshold value $s$. Though the non-directed and non-weighted algorithm already provides convincing results retrieving relevant overlapping set of terms.

To recover the asymmetric aspect of our
paradigmatic measure, we then processed the output of this k-clique analysis by reintroducing the distance between words
within a cluster $C$ and defining two quantities characterizing a
word $w$ within this cluster : the genericity index  $i_g$ and
the specificity index $i_s$ defined as follows:
\begin{description}
  \item[\textsl{The specificity index}] provides the extent to which the word $w$ is specific to the
  cluster
  $C$ with respect to the paradigmatic
  proximity $\Proxm^{\alpha}$ considered (\textsl{i.e.} is $w$ relevant for the terms in $C$ ?). It is the mean of $w$ ``in-paradigmatic proximity" -
  from term $w'\in C$ to $w$ - and is defined as : $$i_s(w)=\frac{1}{card(C)}\sum_{w'\in
  C}\Proxm^{\alpha}(w',w)$$

  \item[\textsl{The genericity index}] defines to what extent the cluster $C$ is a good neighborhood for the word $w$ with respect to the paradigmatic proximity $\Proxm^{\alpha}$. It is the mean of $w$ ``out-paradigmatic proximity" and is defined by : $$i_g(w)=\frac{1}{card(C)}\sum_{w'\in
  C}\Proxm^{\alpha}(w,w')$$
\end{description}

These two indexes enable to plot an intuitive 2D embedding of a
cluster assigning to each term the coordinate $(i_s,i_g)$ and a sphere which size represents the number of its co-occurrences with all the other terms within the cluster and which color is associated to the growth rate of the use of the term within the cluster between two consecutive periods.

\begin{figure}
\centering
  \includegraphics[width=0.9\textwidth]{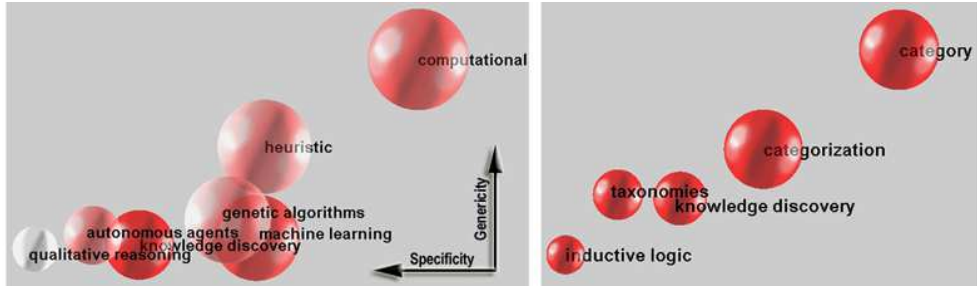}\
  \caption{\textbf{Two paradigmatic fields mentioning the term \textit{Knowledge Discovery} during the period 2002-2005}.
   \textit{Knowledge Discovery} belongs to several
  spheres of knowledge production. Here we can see  one \textit{machine learning} oriented (left), the other concerns \textit{categorization} issues (right). Within a cluster, from left to right $i_s$ decreases, from top to down $i_g$ decreases. The size of the spheres maps according to the number of co-occurrences of the associated term with all the other terms in the cluster. The color of spheres represents the use growth rate of the term within the cluster between 1999-2002 and 2002-2005. A full red point means that the term occurrences have increased of at leat
 150\% between these two periods}\label{cfinder}
\end{figure}

To illustrate this, we present here two fields that share the
term \textit{Knowledge discovery} in the period 2003-2005 (cf. fig.
\ref{cfinder}). As mentionned above, this term indeed belongs to
several fields in our terms set, one more \textit{machine learning}
oriented, the other dedicated to issues of \textit{categorization}.

It should be emphasized here that this meso-scale visualization is
complementary but clearly different from micro-scale visualization.
In this case, only neighbors that satisfy global conditions may
be gathered together. Thus the detected fields outline trends in science according
to a given degree of specificity tuned by $\alpha$. Other examples of automatically reconstructed paradigmatic
fields can be found on \href{http://cssociety.org/CSM}{http://cssociety.org/CSM}.

Given this meso-scale approach, it is now possible to draw a global map of
this set of terms related to complex systems to have a
macro-scale overview.

\section{Macro-scale : science mapping}

The next step is now to give an insight into the articulation of the different paradigmatic fields we have identified in the meso-scale approach in order to provide a global view of the scientific landscape defined by our initial set of terms. For a given period of time, we have defined paradigmatic fields as relevant sets of terms. These terms may belong to several paradigmatic fields. A natural procedure is to consider a weighted graph for which nodes are the paradigmatic fields and edges are defined according to the overlap between paradigmatic fields. Since within each paradigmatic fields we can compute for a term its contextual specificity $i_s$ and genericity $i_g$ indexes, one could in principle use these two indexes to compute the overlap between two paradigmatic fields. We will however consider a much simpler solution and consider that the weight of a link between two paradigmatic fields will be equal to the number of terms shared by these fields. $i_s$  and $i_g$ are nevertheless useful to label automatically the paradigmatic fields by their most generic or most specific terms.

Another information worth displaying on this global map is the growth of activity of each paradigmatic field. 
We again chose a straightforward index and defined a paradigmatic activity growth as the average increase of the normalized occurrences of terms in a target field $C$ between a period $T$ and the same length previous period $T_{-}$:
$$A_C=\frac{1}{card(C)}\sum_{i\in C}\frac{p_i^T}{p_i^{T_{-}}}$$
where $p_i^T$ is defined by $p_i^T=\frac{n_i^T}{\sum_{j}n_j^T}$

\begin{figure}
\centering
  \includegraphics[width=0.9\textwidth]{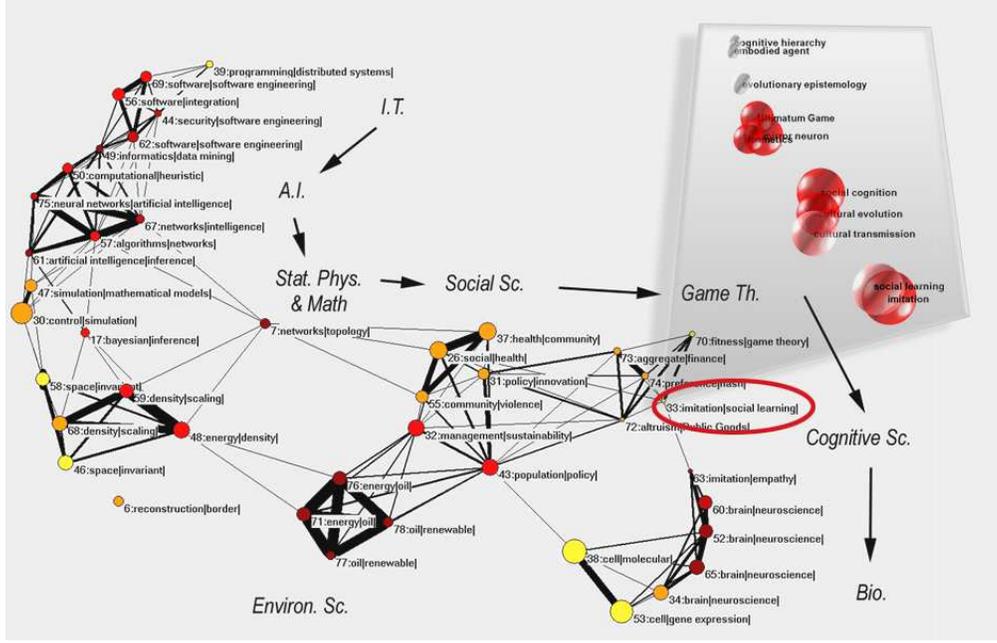}\
  \caption{Macroscopic map of the "Complex Systems" field. The darker the color (from white to deep red) of a paradigmatic field the higher its growth rate. One can zoom into subfields as illustrated in the inset to get further information.}\label{sciencemap}
\end{figure}

The figure \ref{sciencemap} displays the macro-scale map of science on the period 2002-2005 featuring all these information. To keep the overall figure understandable we only plotted the 
 paradigmatic fields that had between 6 and 20 terms.

The size of a node is a logarithmic mapping of the mean normalized number of occurrences of terms in the corresponding cluster and colors indicate the mean activity of the field. Blue colors correspond to a negative growth rate, yellow, red and brown to a positive growth rate. We can notice here that all fields are growing and that the fields are organized in densely connected  areas that can themselves be labeled as large thematic areas namely : biology,  cognitive science, game theory, environmental sciences, social sciences, statistical physics, mathematics, artificial intelligence, computer sciences and information technologies.
An interactive map that allows to zoom in the general map to navigate the paradigmatic fields can be found online: \href{http://cssociety.org/CSM}{http://cssociety.org/CSM}.

\section{Perspective}

We have already sketched methods that provide multi-level description
of our initial set of terms. The next step may be to further study the dynamical properties of  this mutli-level description. Terms occurrences and co-occurrences evolve through time which may trigger sensible change in the way they are structured. If we want to describe the global dynamics of the system at each level, we have to define the dynamics of these meso-pattern which we called paradigmatic fields. Describing this meso-dynamics will require to define a dynamics on these sets of terms featuring gain or loss of some of its constituents, merging of different fields, scission, but also death or birth of new fields. The issue of community dynamics has already been successfully applied to collaboration and cell phone networks \cite{Palla:2007p229} and may happen to be a promising direction of research when applied to science evolution analysis.

\section*{Conclusion}

We tried to show the way science mapping could benefit from an asymmetric proximity measure between terms. Hierarchical and overlapping complex structures have been exhibited and represented in convenient space. Another advantage of this measure is the possibility to perform
multi-scale browsing over a set of terms from the more general to
the more specific, which may happen to be of great help for scientists as well as for other audiences (\textit{e.g. }science policy maker).
Besides the methods exposed are not entirely specific to scientific world ; one can imagine to draw these kinds of knowledge map from any other sources of content (such as patents, press, blogs, etc...).

\section*{Acknowledgements} This study was supported by the CREA - Ecole Polytechnique, the
IST-FET coordinated action ONCE-CS (http://once-cs.csregistry.org) and the Paris
Ile-de France Institute for Complex Systems (http://iscpif.fr).
 We would like to thank Scirus.com for their partnership and Craig Scott
 for his kind help with the data processing, as well as Arc System research for
 their keywords list. 


\end{document}